\documentclass[runningheads]{llncs}

\usepackage[T1]{fontenc}
\usepackage{graphicx}
\usepackage{url}
\usepackage{hyperref}
\hypersetup{
    colorlinks=true,
    linkcolor=blue,
    filecolor=magenta,
    urlcolor=blue,
}
\usepackage{tikz}
\usetikzlibrary{shapes.geometric, arrows.meta, positioning, fit, backgrounds}

\begin{document}

\title{BacPrep: Lessons from Deploying an LLM-Based Bacalaureat Assessment Platform}

\author{Adrian-Marius Dumitran\inst{1,3,4}\orcidID{0009-0005-3547-5772} \and Radu Dita\inst{1}\orcidID{0009-0005-9229-6965} and Angela-Liliana Dumitran \inst{2}\orcidID{0009-0003-3590-9441}} 

\authorrunning{M. Dumitran and R. Dita and A. Dumitran}

% Corrected Institute/Email Section
\institute{University of Bucharest, Faculty of Mathematics and Computer Science, \\ Academiei 14, 010014, Bucharest, Romania\\
\email{marius.dumitran@unibuc.ro, radu.dita@gmail.com} \\ \and
Universitatea Creștină "Dimitrie Cantemir" \\
\email{angela.dumitran@gmail.com}
\and
Cu Drag si Sport SRL, Bucharest, Romania \and Softbinator Technologies, Bucharest, Romania}
\maketitle

\begin{abstract}
Accessing quality preparation and feedback for the Romanian Bacalaureat exam is challenging, particularly for students in remote or underserved areas. This paper presents BacPrep, an experimental online platform exploring Large Language Model (LLM) potential for automated assessment, aiming to offer a free, accessible resource. Using official exam questions from the last 5 years, BacPrep employs the latest available Gemini Flash model (currently Gemini 2.5 Flash, via the \texttt{gemini-flash-latest} endpoint) to prioritize user experience quality during the data collection phase, with model versioning to be locked for subsequent rigorous evaluation. The platform has collected over 100 student solutions across Computer Science and Romanian Language exams, enabling preliminary assessment of LLM grading quality. This revealed several significant challenges: grading inconsistency across multiple runs, arithmetic errors when aggregating fractional scores, performance degradation under large prompt contexts, failure to apply subject-specific rubric weightings, and internal inconsistencies between generated scores and qualitative feedback. These findings motivate a redesigned architecture featuring subject-level prompt decomposition, specialized per-subject graders, and a median-selection strategy across multiple runs. Expert validation against human-graded solutions remains the critical next step.

\end{abstract}

\section{Introduction}
The Romanian Bacalaureat ("Bac") exam is a critical educational milestone, yet 
equitable access to preparation resources, especially personalized feedback, 
remains a challenge, particularly affecting students in remote areas or those 
facing economic hardship. Traditional study methods lack immediacy. The rapid 
evolution of Large Language Models (LLMs) \cite{openai2023gpt4,anthropic2023claude} 
offers opportunities to investigate novel, technology-driven solutions to 
democratize access.

This paper presents BacPrep, an operational experimental platform investigating 
LLM use for automated feedback on Bac practice solutions. Its goals are:
\begin{enumerate}
    \item To provide a free, accessible practice tool using official past exams, 
    delivering experimental LLM-based feedback guided by official grading schemes.
    \item To function as a research testbed, systematically collecting student 
    solutions and evaluating LLM assessment reliability against expert human graders.
\end{enumerate}

Since deployment, the platform has collected over 100 student solutions across 
Computer Science and Romanian Language exams. Preliminary evaluation revealed 
several failure modes in the current assessment approach, motivating a redesigned 
architecture described in this paper. Expert validation against human-graded 
solutions remains the critical next step.

\section{Related Work}

\textbf{Intelligent Tutoring Systems and Automated Assessment.} ITS aim for personalized learning \cite{vanlehn2011relative,ma2014intelligent}, often using automated assessment (AA). AA has evolved for tasks like coding \cite{ihantola2010review} and essay scoring \cite{shermis2013contrasting,attali2006automated}. The complexity of national exams like the Bac challenges traditional AA, motivating LLM exploration.

\textbf{Large Language Models in Education.} LLMs like GPT-4 \cite{openai2023gpt4}, Claude \cite{anthropic2023claude}, and Google's Gemini family show promise for educational tasks \cite{cheng2022gpt,wang2023large}, including assessment \cite{liang2022holistic}. Recent work has demonstrated rapid capability gains on domain-specific CS exams, with leading models advancing from failing grades to near top-student performance within months \cite{dumitran2025struggle}, motivating continued exploration of LLMs for subject-specific automated assessment. However, concerns about reliability, bias, consistency, and feedback quality persist \cite{bommasani2021opportunities}. User acceptance is also crucial \cite{salloum2019factors}.

\textbf{Educational Technology in Romania.} Technology initiatives in Romania often target resource disparities \cite{istrate2017digital}. Automated feedback platforms for Bacalaureat preparation are scarce, making BacPrep one of the first experimental deployments of LLM-based assessment in this specific national exam context.

\section{Platform Design and Methodology}
BacPrep is an operational experimental platform for practice and research data acquisition.

% (Section 3.1 Data Source remains the same)
\subsection{Data Source and Subject Coverage}
The platform utilizes a structured database of:
\begin{itemize}
    \item \textbf{Source}: Official Romanian Bacalaureat exams and models (Ministry of Education).
    \item \textbf{Timeframe}: Past five academic years (approx. 2020-2025).
    \item \textbf{Content}: Questions, associated materials, and official grading schemes ('bareme').
    \item \textbf{Subjects Covered}: Romanian Language \& Literature and Computer Science.
    \item \textbf{Rationale for Focus}: Concentration facilitates expert grader access for validation and aims for deeper data per topic. Adherence to official materials ensures consistency.
\end{itemize}

\subsection{LLM Integration and Assessment Mechanism}
The platform leverages Google's latest generation of Gemini models:
\begin{itemize}
    \item \textbf{LLM Choice}: Employs the latest available Gemini Flash model, accessed via the \texttt{gemini-flash-latest} API endpoint, currently resolving to Gemini 2.5 Flash.
    \item \textbf{Rationale for Choice}: Gemini Flash offers strong quality, large context window, fast response times, and sufficient free RPM for our platform. During the data collection phase we deliberately prioritize user experience by always using the latest available model, with versioning to be locked for subsequent rigorous evaluation.
    \item \textbf{Process}: On submission, a prompt containing the question, student solution, and official grading scheme is sent to the API, instructing strict evaluation against the scheme.
    \item \textbf{Output}: The LLM response is presented to the user (clearly marked as experimental) and logged for research alongside the student solution.
    \item \textbf{Remark}: While the live platform uses \texttt{gemini-flash-latest}, the planned validation study will systematically compare multiple models offline — both proprietary (Gemini, GPT, Claude, Mistral) and open-source (Llama, Gemma, Deepseek, Kimi) — against expert human grades.
\end{itemize}

\subsection{User Interface and Workflow}
To better understand the user experience on BacPrep, we include a walk-through of the main interaction flow a student follows when using the platform.
\begin{figure}[h] \centering \includegraphics[width=0.7\textwidth]{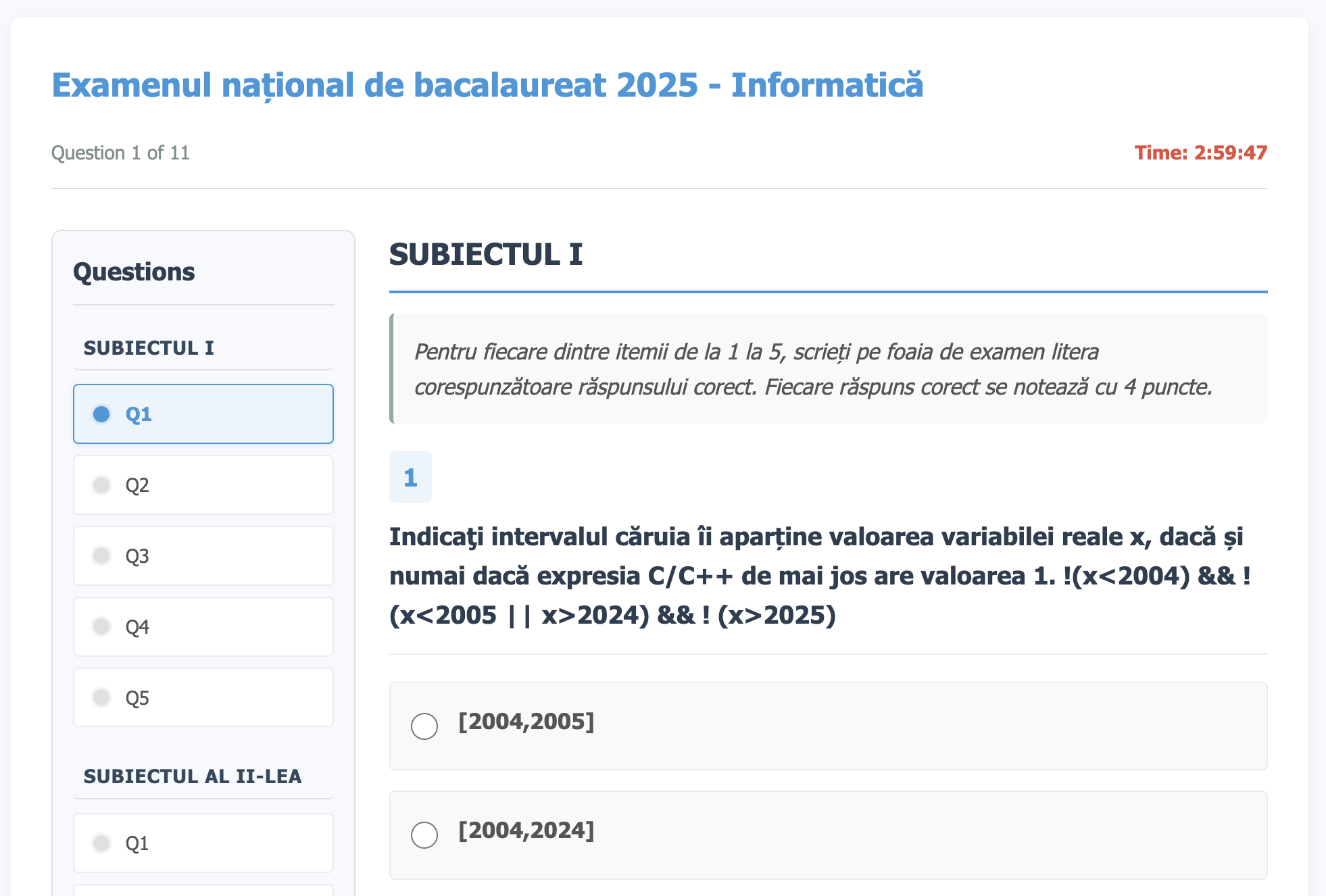} \caption{Example of a multiple-choice Computer Science question in SUBIECTUL I.} \label{fig:question-screen} \end{figure}
\subsubsection{Taking the Exam:} After starting the exam, the student is presented with a structured interface showing grouped questions (e.g., \textbf{SUBIECTUL I}, \textbf{SUBIECTUL AL II-LEA}). Each question has a single or multiple-choice input (Figure~\ref{fig:question-screen}). A timer is displayed for pacing.

\subsubsection{Submission and Evaluation:} 
\begin{figure}[h] \centering \includegraphics[width=0.8\textwidth]{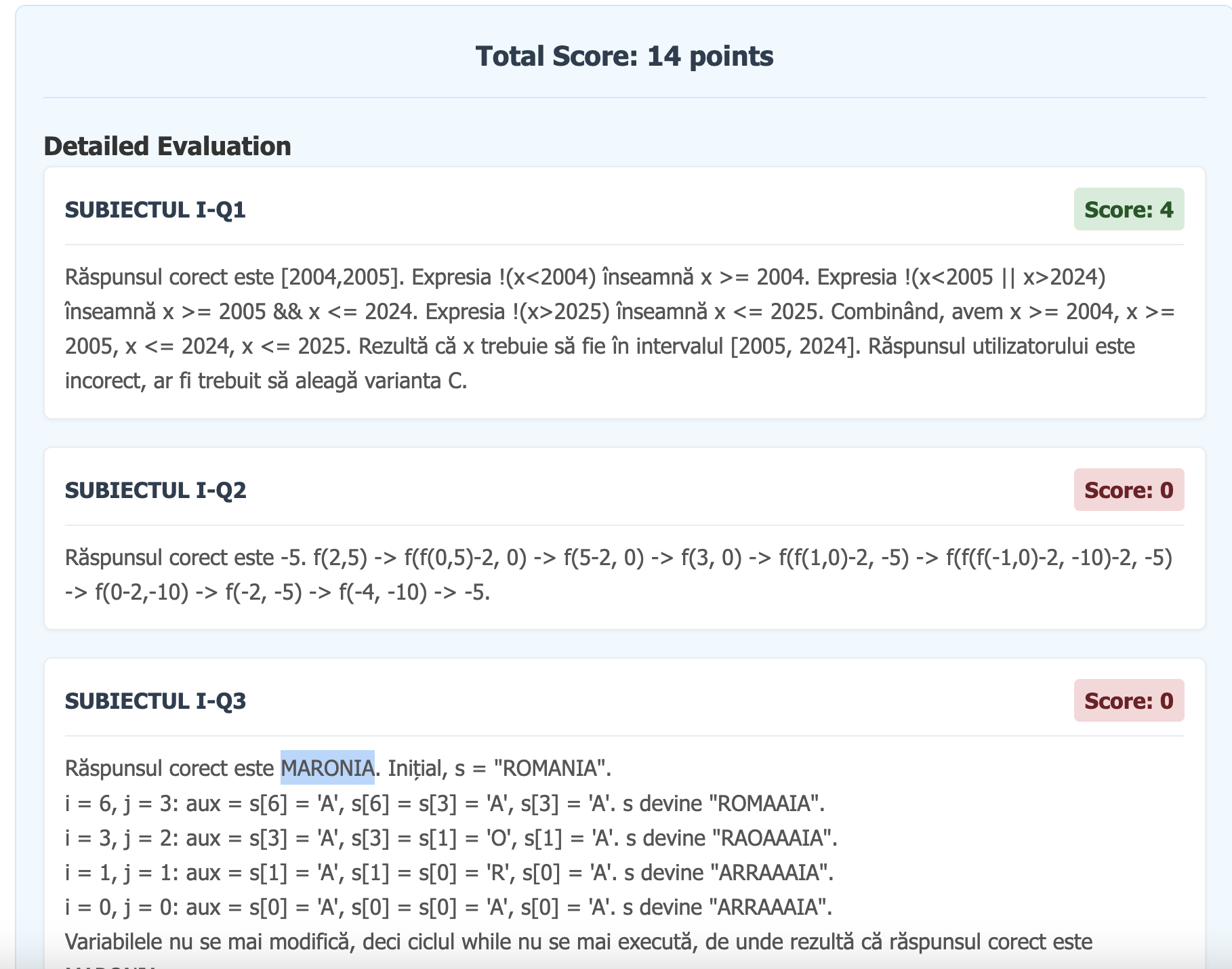} \caption{Automated evaluation of programming responses with output and explanation.} \label{fig:eval-code} \end{figure}

After completing the test, students are shown their score along with a breakdown of each response. The explanation includes reasoning for the correct answer, often accompanied by code evaluations, analysis, or step-by-step deduction (Figure~\ref{fig:eval-code}).

\subsubsection{Session Resume and Progress Tracking:} If a student exits the platform, their exam can be resumed later using the same email. The platform maintains state locally to support continuity in preparation.

\subsection{Ongoing Data Collection}
The platform has collected over 100 student solutions across Computer Science and Romanian Language exams, each paired with the corresponding question and official grading scheme. LLM-generated feedback is logged as metadata alongside each submission. Participation is encouraged through controlled outreach, with 
informed consent covering the experimental nature of the feedback, research goals, and data anonymization.

\section{Preliminary Findings and Redesigned Architecture}
\label{sec:findings}
Following deployment and initial data collection, we conducted a preliminary evaluation of the live assessment system, combining multiple runs on collected student solutions with qualitative feedback from a domain expert. Figure~\ref{fig:architecture} illustrates the redesigned pipeline that directly addresses the systematic failure modes identified, each described below.
While the diagram illustrates the general pipeline, each subject grader is specialized through tailored system prompts reflecting the distinct structure of that subject. For Computer Science exams, graders focus on algorithmic correctness and code evaluation, while Romanian Language graders for Subject III explicitly score formal qualities — grammar, expression, and compositional structure — separately from content, in direct response to initial findings.

\begin{figure}[h]
\centering
\begin{tikzpicture}[
    node distance=0.6cm and 0.8cm,
    box/.style={rectangle, rounded corners, draw=black, fill=white, text width=2.8cm, align=center, minimum height=0.8cm, font=\small},
    grader/.style={rectangle, rounded corners, draw=black!70, fill=gray!10, text width=2.5cm, align=center, minimum height=0.8cm, font=\small},
    arrow/.style={-{Stealth}, thick},
    dasharrow/.style={-{Stealth}, thick, dashed},
    label/.style={font=\footnotesize\itshape, text=gray}
]
\node[box, fill=blue!10] (input) {Full Exam Input};
\node[box, fill=blue!15, below=of input] (decomp) {Subject-Level\\Decomposition};
\draw[arrow] (input) -- (decomp);
\node[grader, below left=0.8cm and 1.6cm of decomp] (s1) {Subject I\\Grader};
\node[grader, below=0.8cm of decomp] (s2) {Subject II\\Grader};
\node[grader, below right=0.8cm and 1.6cm of decomp] (s3) {Subject III\\Grader};
\draw[arrow] (decomp) -- (s1);
\draw[arrow] (decomp) -- (s2);
\draw[arrow] (decomp) -- (s3);
\node[box, fill=orange!10, below=0.8cm of s1] (r1) {Run 1, 2, 3\\Median Selection};
\node[box, fill=orange!10, below=0.8cm of s2] (r2) {Run 1, 2, 3\\Median Selection};
\node[box, fill=orange!10, below=0.8cm of s3] (r3) {Run 1, 2, 3\\Median Selection};
\draw[arrow] (s1) -- (r1);
\draw[arrow] (s2) -- (r2);
\draw[arrow] (s3) -- (r3);
\node[box, fill=green!10, below right=0.8cm and 1.6cm of r1] (agg) {Deterministic\\Score Aggregation};
\draw[arrow] (r1) -- (agg);
\draw[arrow] (r2) -- (agg);
\draw[arrow] (r3) -- (agg);
\node[box, fill=green!15, below=0.8cm of agg] (output) {Final Score +\\Feedback};
\draw[arrow] (agg) -- (output);
\end{tikzpicture}
\caption{Redesigned BacPrep assessment pipeline.}
\label{fig:architecture}
\end{figure}

\begin{itemize}
    \item \textbf{Grading Inconsistency}: Score variance of up to 4 points across multiple runs on the same solution, indicating significant non-determinism even at low temperature settings. $\rightarrow$ \textit{The redesigned system grades each subject multiple times and selects the run closest to the median score.}
    \newpage
    \item \textbf{Arithmetic Errors in Score Aggregation}: Systematic errors when aggregating 15+ fractional sub-scores into a final grade, a well-documented LLM weakness on multi-step numerical tasks. $\rightarrow$ \textit{Score aggregation is delegated to a deterministic tool rather than the LLM itself.}
    \item \textbf{Context Overload}: Assessment quality degrades when the full exam, all solutions, and complete grading scheme are provided in a single prompt. $\rightarrow$ \textit{The redesigned system decomposes the exam by subject, grading each independently.}
    \item \textbf{Rubric Weighting Failure}: For language essays, the model ignored formal qualities (grammar, expression, structure) despite explicit barem allocations. $\rightarrow$ \textit{Specialized per-subject graders encode subject-specific rubric requirements for both content and form.}
\end{itemize}

\section{Validation Strategy}
\label{sec:validation}
The collected student solutions form the foundation of the planned validation strategy, which aims to create a benchmark for evaluating various LLMs on the Bacalaureat assessment task:
\begin{enumerate}
    \item \textbf{Expert Human Grading}: Experienced teachers will grade the collected student solutions using official grading schemes, establishing an expert-verified ground truth dataset.
    \item \textbf{Offline LLM Evaluation}: Using the stored questions, grading schemes, and student solutions, we will systematically query multiple LLMs offline — both proprietary (Gemini, GPT, Claude, Mistral) and open-source (Llama, Gemma, Deepseek) — to generate assessments for each solution.
    \item \textbf{Comparative Analysis}: LLM-generated assessments will be compared against expert grades using quantitative metrics (agreement scores, error analysis) and qualitative review, evaluating accuracy, consistency, and failure modes across models and prompting strategies.
\end{enumerate}

\section{Ethical Considerations}
Operating BacPrep ethically requires continuous attention to three key areas:
\begin{itemize}
    \item \textbf{Managing Feedback Expectations}: Live feedback is clearly marked as experimental via UI disclaimers to prevent over-reliance.
    \item \textbf{Equity vs. Limitations}: The potential equity benefits are balanced with full transparency about the current unproven reliability of automated assessment.
    \item \textbf{Data Privacy}: Student solutions are anonymized, personal data collection is minimal, and all data handling complies with GDPR.
\end{itemize}

\section{Conclusion}
BacPrep is an operational experimental platform investigating LLM potential for 
Bacalaureat assessment while collecting a valuable dataset of authentic student 
solutions. Since deployment, the platform has gathered over 100 student solutions 
across Computer Science and Romanian Language exams. Preliminary evaluation of 
the live assessment system revealed significant failure modes in the single-pass, 
full-exam prompting approach, including grading inconsistency, arithmetic errors, 
context overload, and failure to apply subject-specific rubric weightings. These 
findings directly motivated a redesigned modular architecture featuring 
subject-level decomposition, specialized per-subject graders, multiple runs with 
median selection, and deterministic score aggregation. The immediate next step 
is expert validation — experienced teachers grading the collected solutions to 
establish ground truth — enabling rigorous offline comparison of multiple LLMs 
against human grades. BacPrep serves as an essential testbed for advancing 
empirical understanding of LLM capabilities in high-stakes national exam assessment.

\subsubsection{\discintname} The authors declare no competing interests.

\end{document}